\documentclass[10pt,leqno,titlepage]{amsart}
\usepackage[foot]{amsaddr}
\usepackage{amsmath}
\usepackage{graphicx,psfrag,epsf}
\usepackage{enumerate}
\usepackage{natbib}
\usepackage{url} %
\usepackage{mathtools, empheq}
\usepackage{amssymb, setspace}
\usepackage{bm, bbm}
\usepackage{graphicx, tabularx, array}
\usepackage[svgnames,table]{xcolor}
\usepackage{enumitem}
\usepackage{multirow}
\usepackage{hyperref}
\usepackage{geometry}
\usepackage{makecell}
\usepackage{booktabs}
\usepackage{amsmath}
\usepackage{array}
\usepackage{algorithm}
\usepackage{algorithmic}
\usepackage[bottom]{footmisc} %
\usepackage{longtable}
\usepackage{threeparttable} %
\usepackage{colortbl} %

\newcolumntype{C}[1]{>{\centering\arraybackslash}m{#1}}

\numberwithin{equation}{section}

\usepackage{amssymb,amsthm,amsmath}
\usepackage{etoolbox}
\AtBeginDocument{%
    \let\orignewpage\newpage 
    \renewcommand\newpage{}
    \patchcmd{\clearpage}{\newpage}{\orignewpage}{}{}}

\makeatletter
\def\@settitle{\begin{center}\normalfont\Large\bfseries \@title\end{center}}
\makeatother
\makeatletter
\def\@setauthors{%
  \begingroup
    \let\MakeUppercase\relax
    \begin{center}
        \vspace{1em}
      \normalfont\normalsize\authors
    \end{center}%
  \endgroup
}
\makeatother

\begin{document}
\bibliographystyle{plainnat}

\def\spacingset#1{\renewcommand{\baselinestretch}%
{#1}\small\normalsize} \spacingset{1}

\newlist{steps}{enumerate}{1}
\setlist[steps, 1]{label = Step \arabic*:}

\title[]{Protocol for an Observational Study on the Effects of Adolescent Physical Activity and Sports Participation on Flourishing and Academic Engagement}

\author{William Bekerman$^{1}$} \email{bekerman@wharton.upenn.edu}
\author{Rebecca E. Hasson$^{2}$}
\email{hassonr@umich.edu}
\author{Leah E. Robinson$^{2}$}
\email{lerobin@umich.edu}
\author{Dylan S. Small$^{1}$}
\email{dsmall@wharton.upenn.edu}

\dedicatory{$^{1}$Department of Statistics and Data Science, University of Pennsylvania, Philadelphia, PA, USA\\$^{2}$School of Kinesiology, University of Michigan, Ann Arbor, MI, USA}

\begin{abstract}
The wide-ranging benefits of physical activity and sports participation among children and adolescents have been closely examined and are reported to include improvements in physical and mental health, cognitive functioning, and social connectedness. However, it remains largely unknown how these activities may affect flourishing and academic engagement, which are closely tied to long-term success, health, and well-being, and how these patterns evolved before, during, and after the COVID-19 pandemic. In this article, we provide the protocol for an observational study using data from the National Survey of Children’s Health to examine these relationships among American adolescents. To strengthen our findings, we will conduct this investigation across three different time periods, allowing us to assess the replicability of our conclusions. We introduce a novel statistical design, called data turnover, to carry out this analysis. Data turnover allows a single group of statisticians and domain experts to work together to assess the strength of evidence gathered across multiple data splits while incorporating both qualitative and quantitative findings from data exploration. We delineate our analysis plan using this new method and conclude with a brief discussion of additional considerations for our study.
\end{abstract}

\maketitle

\spacingset{1.5}

\section{Background and Motivation}
Physical activity and sports participation have long been associated with favorable physical health in children and adolescents, with well-documented beneficial effects on adiposity, musculoskeletal health and fitness, cardiovascular health, and blood pressure \citep{janssen2010systematic}. Yet, physical activity and sports participation may also contribute to positive developmental outcomes beyond physical health. For example, there has been some evidence linking it to flourishing and academic engagement in adolescents \citep{owen2016physical,dyer2017sport,zhou2026physical}. Meanwhile, it has been suggested that discontinuing sports participation may be negatively associated with academic performance \citep{ishihara2020relationship}. Flourishing encompasses characteristics such as curiosity in learning, task persistence, and self-regulation, which may have important implications for long-term health and success \citep{duckworth2007grit,bub2016longitudinal}. Likewise, school engagement is closely related to improved academic achievement and subjective well-being \citep{wong2024student}. 

Notably, youth activity levels have decreased markedly in recent decades \citep{boreham2001physical} due to factors such as increased daily screen time and consumption of digital media, shorter recess times, and fear of bullying. Indeed, over 80\% of adolescents worldwide are insufficiently active, posing a substantial public health challenge \citep{guthold2020global}. Despite some studies pointing to links between physical activity and flourishing and academic outcomes, these patterns are not fully clear. Moreover, it is unknown how these connections were affected by the significant societal disruptions that occurred due to the COVID-19 pandemic in early 2020. Importantly, containment strategies like social distancing and school closures negatively impacted opportunities for organized sports and activity, which typically resulted in notable decreases in physical activity levels \citep{Wunsch2022changes}. We look to clarify how physical activity and sports participation are related to flourishing and academic engagement among American adolescents and whether these relationships changed before, during, and after the COVID-19 pandemic.

We will examine these relationships using a large national sample from the National Survey of Children’s Health (NSCH), a detailed household survey that provides national and state-level data on the health and well-being of American children and adolescents in the United States (see, for instance, \cite{van2004national}). We will focus on adolescents ages 12 to 17 and jointly examine these patterns across multiple eras of data collection in order to assess both their evolution and replicability. Importantly, the large sample size of the NSCH (each year of data has over $20,000$ respondents) will help facilitate powerful data analysis and may lead to new discoveries that extend the current literature. The NSCH also collected detailed measurements of variables that may confound the relationships between activity and our outcomes of interest, such as those related to personal demographics and family background. For instance, \cite{bekerman2026powerful} identified significant differences in vigorous physical activity between adolescent girls below and above the poverty line, and previous literature has noted that variables like race/ethnicity and gender may confound these relationships. We will control for these variables in our regression models to lessen the burden of confounding biases. 

We introduce a novel statistical design, called \textit{data turnover}, to conduct this analysis. Data turnover allows for a single group of statisticians and domain experts to work together to assess the strength of evidence gathered across multiple data splits, while incorporating both qualitative and quantitative findings from data exploration. Exploratory data analysis allows us to look at our data before making any (potentially dubious) assumptions, while enabling the recognition of any obvious errors, detection of outliers or anomalies, and shaping of new hypotheses. It also permits us to identify a variable that does not measure what we initially thought and help suggest a new one, or facilitate data-informed hypothesis selection (\cite{bekerman2026planning}). Along with exploration, replication is a valuable tool for statisticians conducting observational studies. Replication of scientific results lends credence to these findings and serves to maintain and build trust in scientific knowledge as a whole. Previous works have facilitated analyses that allow for exploration but no replication (e.g., \cite{cox1975note,heller2009split}), replication but no exploration (e.g., \cite{zhao2018cross, karmakar2019integrating}), or both under the stringent requirement of two independent teams of investigators (\cite{roy2022protocol}). Data turnover allows us to explore the data and assess the replicability of our findings with only one group of researchers.

Evaluating the replicability of our findings will be made possible by analyzing the data separately from respondents across three different time periods -- those surveyed pre-COVID (2016 - 2019), during COVID (2020 - 2021), and post-COVID (2023 - 2024). In observational studies like ours where we do not control assignment of individuals to receive treatment, obtaining similar results across multiple groups with different treatment assignment mechanisms strengthens the evidence that the treatment is in fact the cause of its ostensible effects (\cite{rosenbaum2001replicating,rosenbaum2015see}). Paul Rosenbaum (\citeyear{rosenbaum2015cochran}) would liken our approach to accumulating evidence in a crossword puzzle: while each piece of evidence may be tentative when considered individually, assessing them jointly can lead to more convincing conclusions.  
We refer to those hypotheses which are significant across at least two time periods as ``replicable'' findings (\cite{bogomolov2023replicability}). In Section \ref{analysis}, we provide a detailed description of how we use this non-random cross-screening approach to identify replicable hypotheses. We also discuss testing of the global null hypothesis; that is, identifying those hypotheses which are deemed significant in at least one of the three time periods.

The remainder of the document is organized as follows. We introduce the NSCH dataset and describe our data processing in Section \ref{NSCH}. In Section \ref{covariates-list}, we discuss the covariates we intend to control for in our analysis, and report our primary exposures and outcomes of interest in Section \ref{sec:outcomes}. Finally, we conclude with a description of our data analysis procedure in Section \ref{analysis}. %

\section{National Survey of Children’s Health (NSCH) Dataset} \label{NSCH}

The NSCH is an extensive cross-sectional survey on the health and well-being of American youth. Detailed information pertaining to family background,  education, health, economic status, and much more, were collected across various surveys between 2003 and 2024.

To investigate the relationships of physical activity and sports participation with flourishing and academic engagement, we collect these variables and potential confounding variables for each year between 2016 and 2024, excluding 2022. We look to divide these years into distinct time periods that broadly represent the pre-COVID, COVID, and post-COVID eras, so we choose 2016 - 2019, 2020 - 2021, and 2023 - 2024, respectively. We omit the year 2022 from these time periods since we had difficulty classifying it into either the COVID or post-COVID era. To ensure that our analysis accurately reflects the population of American adolescents, we incorporate the survey weights provided by the NSCH into our descriptive statistics and analysis plan. For ease of conducting our study, we will use any imputations where they are done by the NSCH, but not impute any data on our own. Then, we will conduct a complete-case analysis and consider only those observations with no remaining missing data. We remove those respondents who are missing our variables of interest or any confounding variables. 

Among the remaining subjects, we have a total of 42,248 individuals in the pre-COVID era, 26,123 in the COVID era, and 31,252 in the post-COVID era. These samples are cross-sectional, so individuals observed in each era need not be the same subjects. Among these, 26,704 individuals participated in sports in the pre-COVID era, compared with 14,559 subjects and 18,728 subjects in the COVID and post-COVID eras, respectively. Notably, roughly 87\% of adolescents had at least one day per week exercising at least 60 minutes pre-COVID, which decreased to 84\% during COVID, before rebounding slightly to roughly 86\%. However, less than half of these individuals maintained this activity level for at least four days per week, across all time periods.

\section{Baseline Covariates} \label{covariates-list}

In order to lessen the burden of potential confounding biases, we control for relevant baseline covariates that may confound the relationships between activity and our outcomes of interest. 
We gather variables related to personal demographics and family background to account for potential confounding biases in our data analysis. We list the corresponding variable name in the NSCH dataset in parentheses. These covariates are listed as follows:

\begin{itemize}
    \item Federal poverty level (\texttt{fpl}).
    \item Race/Ethnicity (\texttt{sc\_race\_r}).
    \item Sex (\texttt{sc\_sex}).
    \item Age (\texttt{sc\_age\_years}).
    \item Highest level of parental education (\texttt{higrade\_tvis}).
    \item Neighborhood safety (\texttt{k10q40\_r}).
    \item Family structure (\texttt{family\_r}).
\end{itemize}

We modify the parental education, neighborhood safety, and family structure variables. For parental education, we code the possible responses "Less than High School" as 10, "High School" as 12, "Some College or Associate Degree" as 14, and "College Degree or Higher" as 16, to reflect the number of years typically required to attain the stated education level. To evaluate neighborhood safety, the parent is asked how strongly they agree/disagree that their child is safe in their neighborhood. We code "Definitely agree" as one, and take the remaining responses to be zero. For family structure, we consider two-parent households versus those without two parents. In particular, two-parent households comprise those who responded as having two biological/adoptive parents (married), two biological/adoptive parents (not currently married), two parents (at least one not biological/adoptive, currently married), or two parents (at least one not biological/adoptive, not currently married).

\begin{figure}[bt]
\centering
\includegraphics[width = \textwidth-11em]{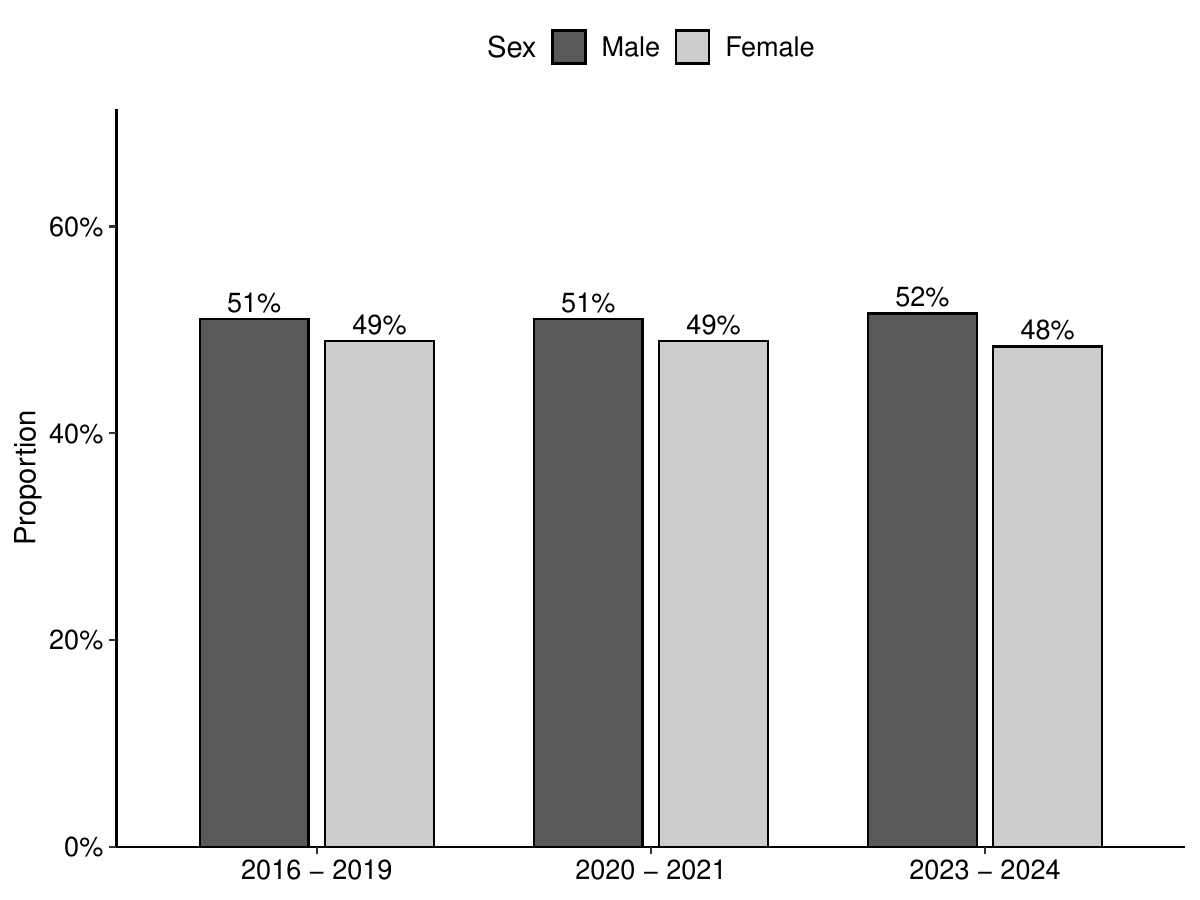}
\caption{Bar plots showing the proportion of the study sample by sex across eras of survey data.}
\label{fig:gender}
\end{figure}

\begin{figure}[b]
\centering
\includegraphics[width = \textwidth-11em]{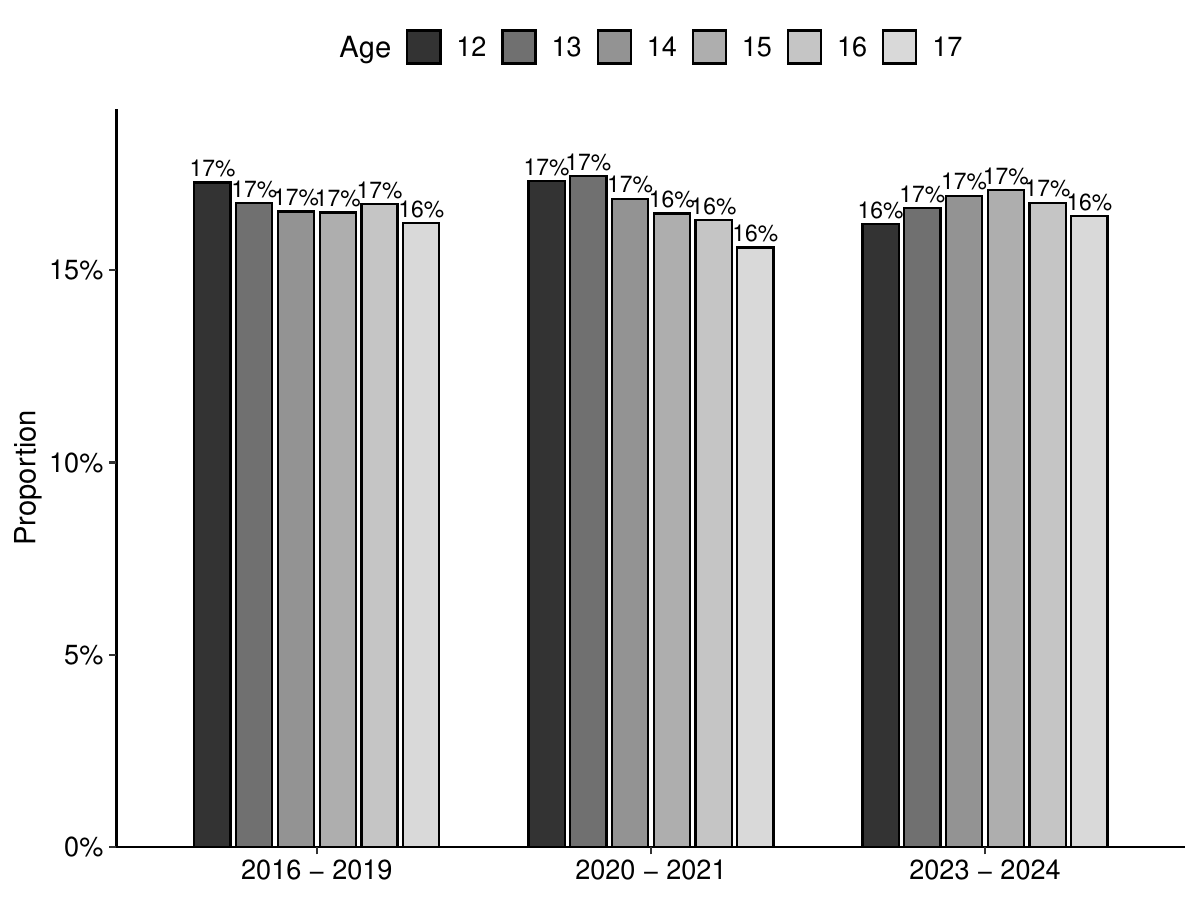}
\caption{Bar plots showing the proportion of the study sample by age across eras of survey data.}
\label{fig:age}
\end{figure}

We show the distributions of sex and age for the remaining survey respondents in Figures~\ref{fig:gender} and \ref{fig:age} below \footnote{All plots presented in this manuscript are descriptive in nature and are generated after data processing. Exact proportions may differ slightly.}. In these plots, we observe that there is very little covariate shift between eras of survey responses. We discuss how we control for these variables and other measured confounders in Section~\ref{analysis}. 

\section{Exposures and  Outcomes of Interest} \label{sec:outcomes}

We collect the exposures and outcomes of interest to our study, listing the corresponding variable name(s) in the NSCH dataset in parentheses. The exposures are given as follows:
\newline \newline \newline
\textbf{Physical activity}
\begin{itemize}   
    \item How many days of exercise, playing a sport, or participating in physical activity for at least 60 minutes during the past week? (\texttt{physactiv}) 
\end{itemize}
\textbf{Sports participation}
\begin{itemize}
    \item {Any participation in a sports team or sports lessons after school or on weekends during the past 12 months? (\texttt{k7q30})}
\end{itemize}
We note that, for the physical activity exposure, we modify the coding such that zero days is coded as 0, between one and three days is coded as 2, between four and six days is coded as 5, and every day is coded as 7. We do not alter the sports participation variable. Next, we list the pre-specified outcomes under study:
\newline
\textbf{Flourishing}
\begin{itemize}
    \item How often does the child show curiosity in learning new things, work to finish tasks they start, and stay calm when faced with a challenge? (\texttt{k6q71\_r, k7q84\_r, k7q85\_r})
\end{itemize}
\textbf{Academic engagement}
\begin{itemize}
    \item How often does the child care about doing well in school and do all required homework? (\texttt{k7q82\_r, k7q83\_r})
\end{itemize}
The flourishing outcome is the sum of the three individual items listed above, where responses of "Always" or "Usually" are coded as one and all other responses as zero. The flourishing index is therefore a score between zero and three. We define the academic engagement outcome as the sum of its two individual items, where an answer of ``Always'' yields three points, ``Usually'' two points, ``Sometimes'' one point, and ``Never'' zero points. Thus, the school engagement composite score can range between zero and six.

\begin{figure}[t]
\centering
\includegraphics[width = \textwidth-11em]{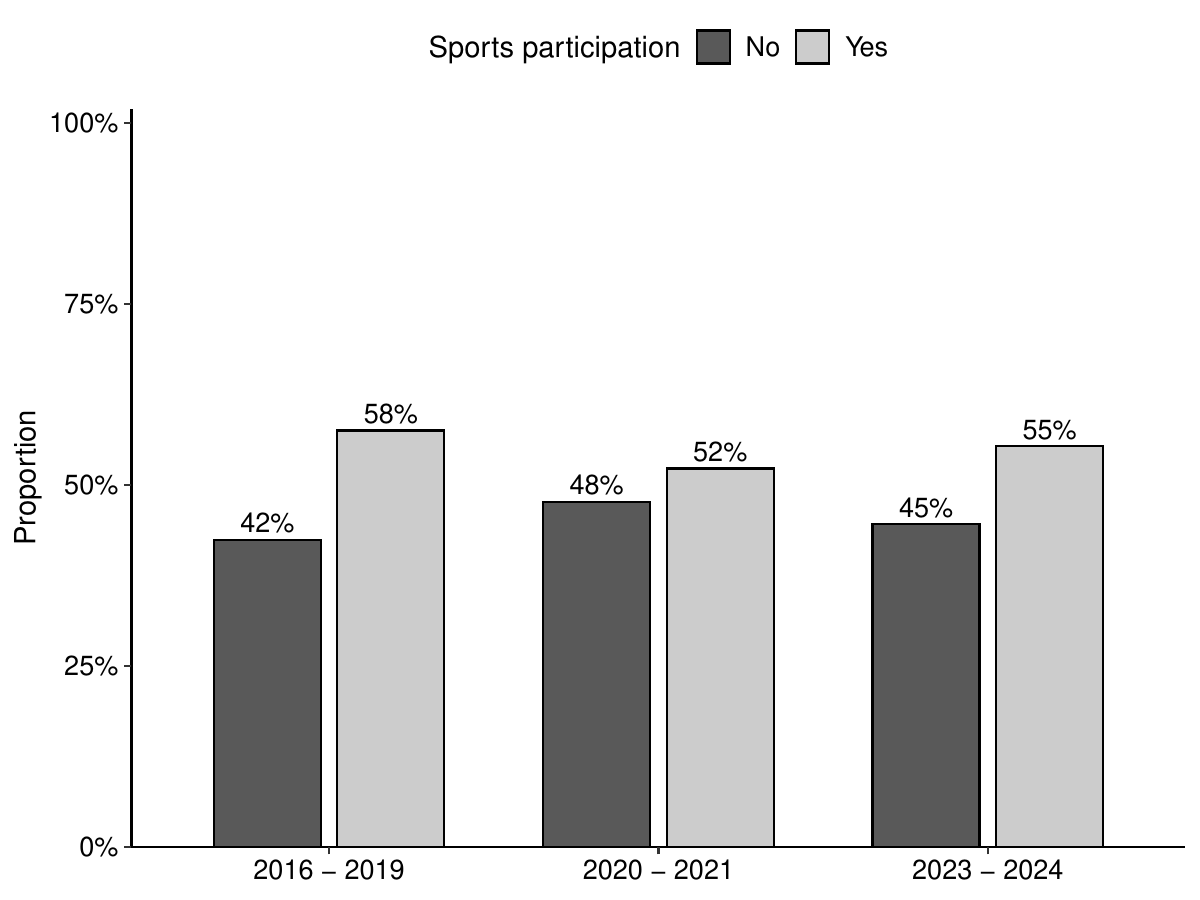}
\caption{Bar plots showing the proportion of the study sample by sports participation across eras of survey data}
\label{fig:sports}
\end{figure}

\begin{figure}[b]
\centering
\includegraphics[width = \textwidth-11em]{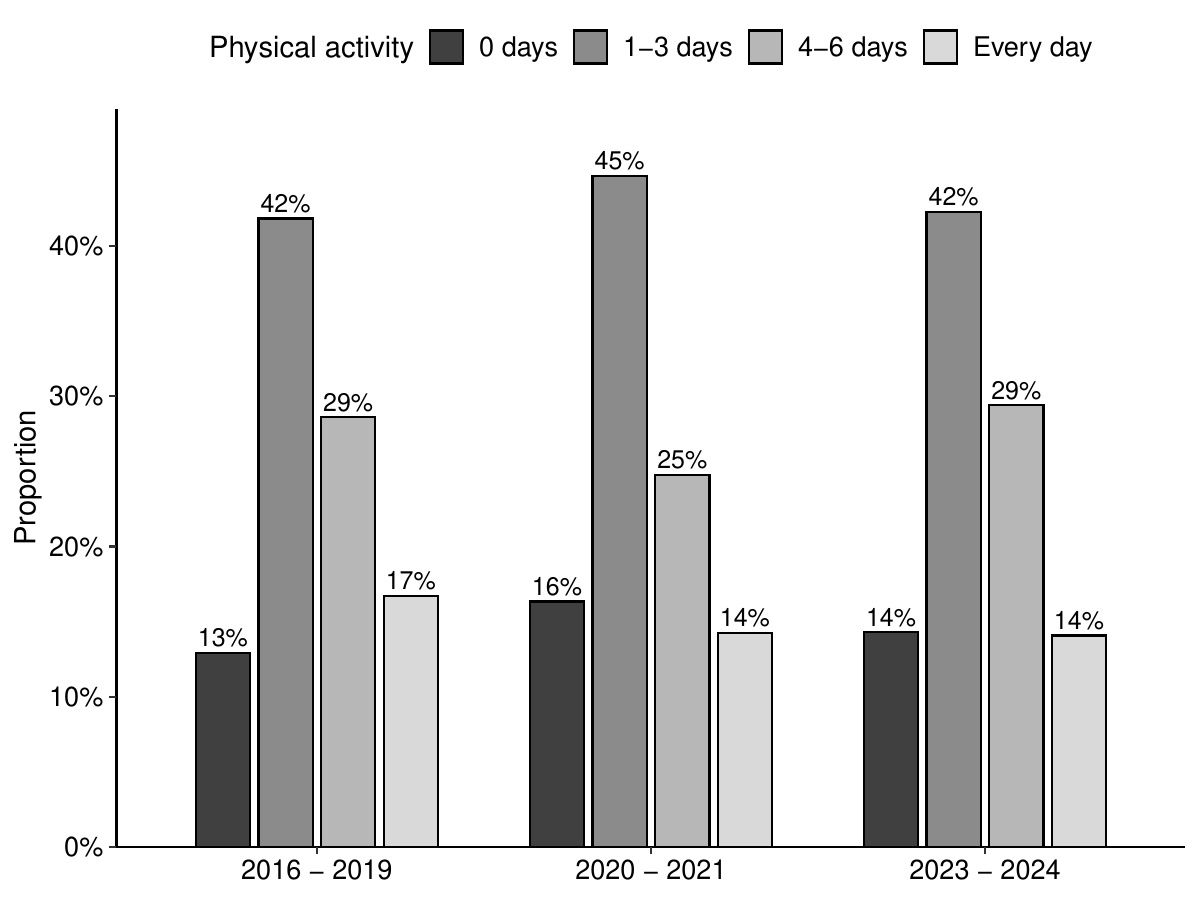}
\caption{Bar plots showing the proportion of the study sample by physical activity across eras of survey data}
\label{fig:activity}
\end{figure}

\begin{figure}[t]
\centering
\includegraphics[width = \textwidth-11em]{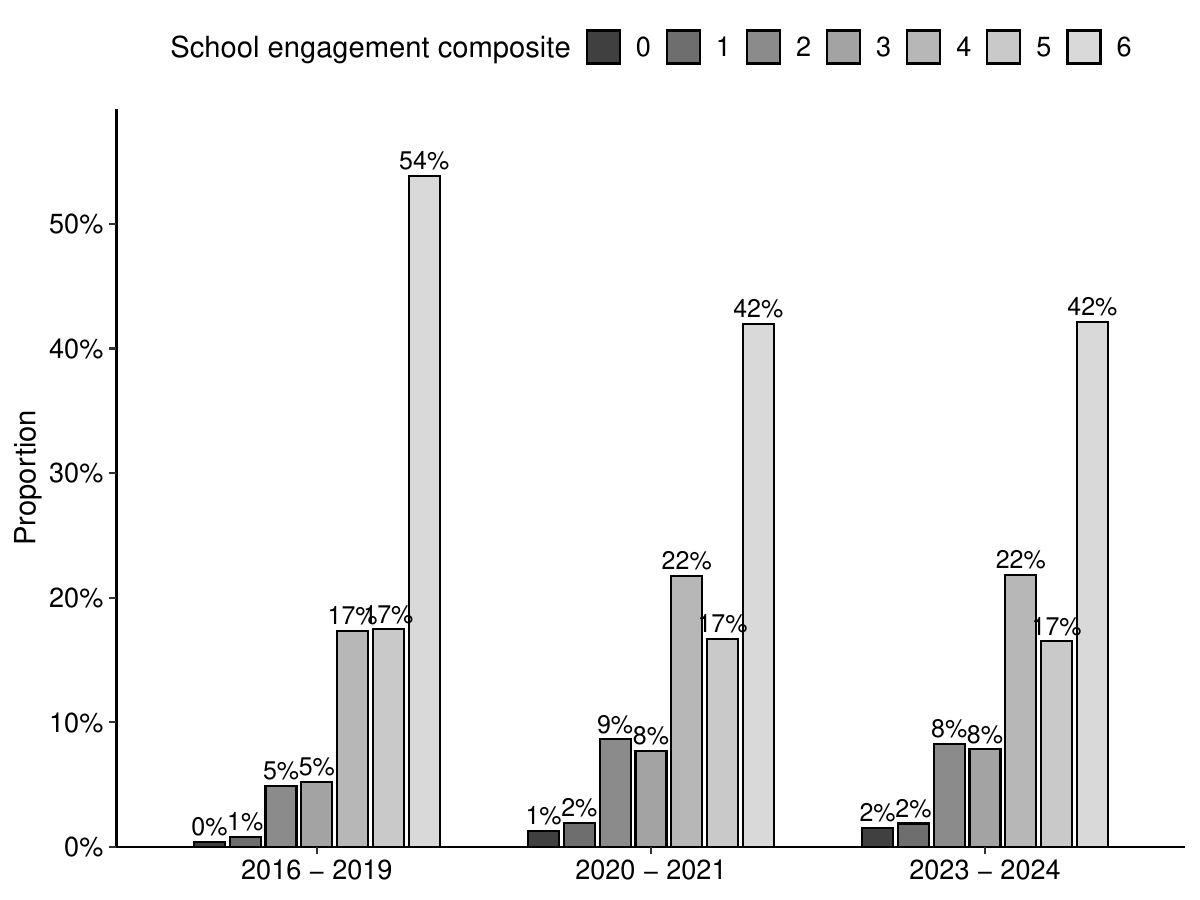}
\caption{Bar plots showing the proportion of the study sample by school engagement across eras of survey data}
\label{fig:school}
\end{figure}

\begin{figure}[b]
\centering
\includegraphics[width = \textwidth-11em]{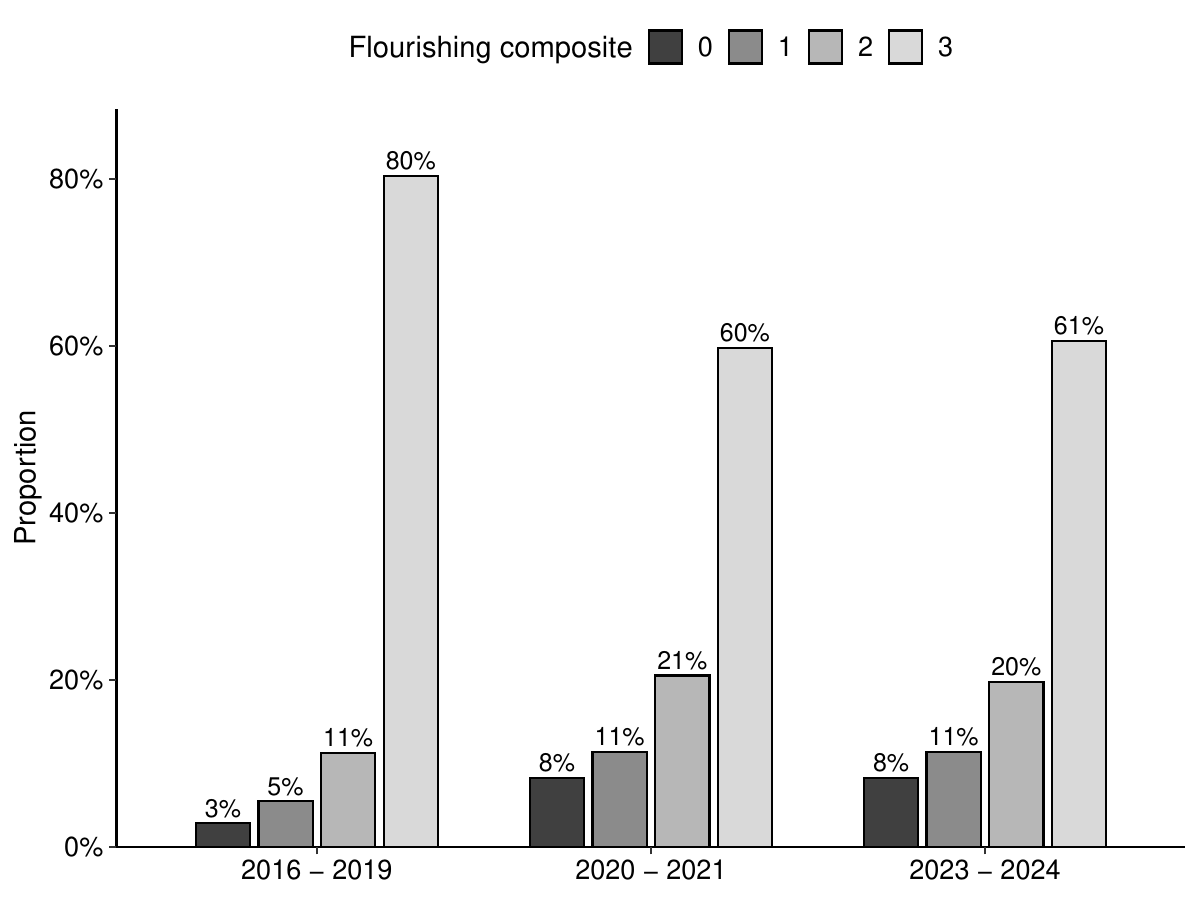}
\caption{Bar plots showing the proportion of the study sample by flourishing across eras of survey data}
\label{fig:flourishing}
\end{figure}

We visualize the distributions of sports participation and physical activity in Figures~\ref{fig:sports} and \ref{fig:activity} provided below. We observe that more adolescents tend to participate in sports, although COVID appeared to negatively impact this involvement. Additionally, we see that a large proportion of adolescents obtain between one and six days of physical activity, with relatively few not exercising at all over the past week. We also depict the school engagement and flourishing outcomes in Figures~\ref{fig:school} and \ref{fig:flourishing}. Most parents provided favorable responses to these items, and both distributions appear notably left-skewed. Yet the proportion of respondents who provided the most favorable responses decreased markedly during COVID and did not rebound.

\section{Data Analysis} \label{analysis}

We seek to identify whether physical activity and sports participation are associated with flourishing and school engagement among adolescents, and whether these associations replicate across the three eras of survey data. We will deem those hypotheses for which the effect is found to be significantly negative or positive for at least two eras as \textit{replicable} findings. We will also establish which of the outcomes are affected by the exposures in at least one of the three eras, a weaker yet still informative standard of evidence akin to meta-analysis. We can also examine if any hypotheses are significant across every era, representing the most compelling evidence of association between exposure and outcome.

We outline our overall analysis procedure as follows. We will use a prespecified plan (described below) to test the four exposure/outcome hypotheses on the 2023-2024 data, using information from a prespecified linear regression on the 2016-2019 data to select the alternative hypotheses. We will then explore the 2023-2024 data and choose in any manner how to formally examine the 2016-2019 data. We will carry out the analyses for both the 2016-2019 and 2023-2024 data. Afterwards, we will explore both of these datasets and examine our results to inform our analysis of the 2020-2021 data. This is a generalization of the data turnover method delineated in \cite{bekerman2024protocol}.

We now discuss our plan to analyze the 2023-2024 data. For both outcomes, we will fit linear regression models relating the particular outcome to each exposure of interest, while adjusting for the measured confounders described in Section~\ref{covariates-list}. This will provide us with four separate regression models, which we write as follows:
\begin{align*}
    &\text{Flourishing} = \beta_0 + \beta_1 {\text{Activity}} + \boldsymbol{\gamma}^\top \bm{X} + \varepsilon,\\
    &\text{Flourishing} = \beta_0 + \beta_1 {\text{Sports}} + \boldsymbol{\gamma}^\top \bm{X} + \varepsilon,\\ 
    &\text{Engagement} = \beta_0 + \beta_1 {\text{Activity}} + \boldsymbol{\gamma}^\top \bm{X} + \varepsilon,\\
    &\text{Engagement} = \beta_0 + \beta_1 {\text{Sports}} + \boldsymbol{\gamma}^\top \bm{X} + \varepsilon.
\end{align*}

For each model, we regress the outcome on the particular exposure, measured covariates $\bm X$, and intercept term while accounting for the NSCH survey design and survey weights. Our primary inferential target is each $\beta_1$ term, which quantifies the adjusted association between the particular exposure and outcome in the model. Specifically, we will test for each model the null hypothesis $H_{0}:\beta_{1}=0$ against the one-sided alternative suggested from fitting these same models on the 2016-2019 data. In particular, if $\widehat{\beta}_{1} < 0$ on the pre-COVID data, we will test on the post-COVID data against $H_{1}:\beta_{1}<0,$ and vice-versa if the estimate is positive. Under consistency, positivity, conditional exchangeability given the measured covariates, and correct specification of the conditional mean model, including a linear dose-response for physical activity and a constant additive exposure effect, the exposure coefficient identifies the average causal effect of interest. We will conduct inference using robust $t$-tests based on survey-design sandwich standard errors.

We propose a novel strategy to control the type I error rate of our analysis. Let $\ell(k)\in\{0,1,2,3\}$ be the number of time periods where the null hypothesis $k$ is false. Our procedure outputs for each $k=1,\dots,4$ a lower bound $\hat{\ell}(k)$ on the number of time periods in which $k$ is non-null. A lower bound $\hat{\ell}=0$ is a trivial lower bound, a lower bound $\hat{\ell}=1$ corresponds to the claim that $k$ is non-null in at least one of the time periods, i.e., a claim that the global null is false for this hypothesis. A lower bound $\hat{\ell}=2$ corresponds to the claim that $k$ is non-null in at least two time periods, i.e., a replicability claim, whereas $\hat{\ell}=3$ is the strongest claim we can make for a hypothesis in this study. Our \textit{data turnover} method will guarantee control of the family-wise error rate (FWER) for these lower bounds, i.e., the probability that for at least one hypothesis $k,$ its lower bound $\hat{\ell}(k)$ exceeds the true number of time periods where the null is false, at no more than $\alpha = 0.05$ (see \cite{benjamini2008screening} and \cite{benjamini2009selective} for similar error measures). Controlling the FWER of each time period at level $\alpha/3$ using Bonferroni correction allows us to maintain $\alpha$-level control of our overall analysis and inferential claims. Our new method may increase detection power compared to automated cross-screening (\cite{zhao2018cross}) by allowing for a data-informed design plan and facilitating qualitative and quantitative insights from exploration, as well as the potential formation of novel hypotheses. 

We expect that incorporating data exploration will enhance the overall design of our study by facilitating data-informed design decisions and accommodating the possible development of new hypotheses. The flexibility afforded by our approach may allow for even further benefits we have not outlined nor yet anticipated. Moreover, the NSCH dataset is richer than the list of covariates and outcomes outlined in Sections \ref{covariates-list} and \ref{sec:outcomes}, so an unrestricted look at the data may facilitate the development of additional novel hypotheses and related outcomes. 

Finally, we are also interested in reporting effect sizes for the rejected null hypotheses with well-defined parameters. Along with forming marginal one-sided confidence intervals for the tested parameters, we also suggest constructing multiplicity-adjusted one-sided confidence intervals within each era so that the collection of intervals attains simultaneous coverage at the corresponding FWER level. If \(r\) parameters are tested in a particular era at level \(\alpha/3\), we will construct confidence intervals for these parameters at level \(1-\alpha/(3r)\).

\bibliography{references.bib}

\end{document}